\documentclass[superscriptaddress,twocolumn,showpacs,prb,amsmath,amssymb]{revtex4}
\usepackage[varg]{txfonts}

\usepackage{graphicx}
\usepackage{bm}

\def\XXint#1#2#3{{\setbox0=\hbox{$#1{#2#3}{\int}$}
    \vcenter{\hbox{$#2#3$}}\kern-.5\wd0}}

\begin{document}

\title{Mesoscopic gap fluctuations in an unconventional superconductor}

\author{Victor~Galitski}
\affiliation{Department of Physics and Joint Quantum Institute,
University of Maryland, College Park, MD 20742-4111}

\begin{abstract}
We study mesoscopic disorder fluctuations in an anisotropic gap
superconductor, which lead to the spatial variations of the local
pairing temperature and  formation of superconducting islands
above the mean-field transition. We derive the probability
distribution function of the pairing temperatures and
superconducting gaps. It is shown that  above the mean-field
transition, a disordered BCS superconductor with an unusual
pairing symmetry is described by a network of superconducting
islands and metallic regions with a strongly suppressed density of
states due to superconducting fluctuations. We argue that the
phenomena associated with mesoscopic disorder fluctuations may
also be relevant to  the high-temperature superconductors, in
particular, to recent STM experiments, where gap inhomogeneities
have been explicitly observed. It is suggested that the gap
fluctuations in the pseudogap phase should be directly related to
the corresponding fluctuations of the pairing
temperature. 
\end{abstract}

 \maketitle Understanding the
phase diagram and the properties of the high-temperature and other
unconventional superconductors has been among the most complex
problems of modern condensed matter physics. Most current
theoretical approaches to the problem concentrate on strong
correlation physics and usually assume that the effects of
disorder are unimportant. However, there exist a number of recent
experimental works, in particular STM studies of the high-$T_c$
cuprates,\cite{RMP.STM,Yazdani,E,E2,E3,E4,E5} which provide a
tentative indication that at least in some materials disorder
plays an important role in the local formation of the
superconducting gap. In particular, Gomes et al.~\cite{Yazdani}
have studied the local development of the gap as a function of
temperature in Bi$_2$Sr$_2$CaCu$_2$O$_{8+\delta}$ above the
superconducting transition and up to a pseudogap temperature,
where the gap inhomogeneities cease to exist. An important result
of this experiment is that the real-space gap map observed was
static and reproducible. This strongly suggests that the
inhomogeneous gap formation is unlikely to be a phase-separation
or superconducting fluctuation effect, but is due to some kind of
disorder in the system.

Motivated by these experiments, we theoretically consider a
disordered superconductor with an unusual pairing symmetry (e.g.,
a $d$-wave superconductor) and study mesoscopic variations of the
local pairing temperature. 
 We point out that the existence of the
Griffiths-type~\cite{Griffiths,Grif_exp,myGrif} phase in the
superconducting phase diagram is specific to an anisotropic gap
superconductor and should not occur in the conventional $s$-wave
systems (due to Anderson theorem), unless they are extremely dirty
or time-reversal symmetry
is broken.\cite{SZ,LS,GaL} 
 An important observation is that if the pairing gap is
anisotropic, the Anderson theorem breaks down and the
superconducting pairing temperature, $T_p$, is suppressed by
disorder even if time-reversal symmetry is preserved (here and
below we make a distinction between the pairing temperature,
$T_p$, and the superconducting transition temperature, $T_c$,
although in the framework of the weak-coupling BCS theory they are
essentially the same).  The impurities are positioned randomly in
space and their density is a random variable. Thus, there always
exist regions where the distribution of impurities is such that
the local pairing temperature, $T_p({\bf r})$, is larger than the
system-wide average value, $\langle{T_p}\rangle$, and the
experimental temperature, $T$.\cite{IL} These regions form islands
with a well-defined gap, which exist on the background of a metal,
if $T_p({\bf r}) > T
> \langle{T_p}\rangle$. The width of the ``mesoscopic fluctuation''
region certainly depends on the strength of disorder and the only
parameter, which may enter this dependence, is the dimensionless
conductance. This defines a narrow window where the
disorder-induced Griffiths phase co-exists with strong
superconducting fluctuations. Therefore,  the picture of impurity
induced inhomogeneities in an anisotropic gap BCS superconductor
is that of superconducting islands and metallic regions with
strongly suppressed density of states.

We start with the following Hamiltonian with a built-in $l$-wave
pairing ($l > 0$):
\begin{eqnarray}
\label{H}
 \hat{\cal H} =\!\!\!\! \int d^2{\bf r}\left\{
\hat\psi^{\dagger}({\bf r}) \left[ - \frac{{\bm \nabla}^2}{2m} -
\mu + U({\bf r}) \right] \hat\psi({\bf r}) -\lambda_l
\hat{b}^\dagger({\bf r}) \hat{b}({\bf r})\right\},
\end{eqnarray}
where $U({\bf r})$ is a disorder potential, $\lambda_l$ is the
$l$-wave interaction constant, and $b({\bf r})$ corresponds to an
``$l$-wave Cooper pair'' $\hat{b}({\bf r}) = \sum\limits_{{\bf
k},{\bf q}} \chi_l(\phi) \hat{\psi}({\bf k}+{\bf q}/2)
\hat{\psi}({\bf p}-{\bf q}/2) e^{i {\bf q}\cdot {\bf r}}$, with
$\phi$ being the angle between the direction  of the vector ${\bf
k}$ and the $x$-axis and  $\chi_l(\phi)$ is the function, which
enforces the $l$-wave symmetry of the gap in the mean-field.

The first step is to integrate out the fermions and express the
action in terms of the order parameter $\Delta_{\bf k} =
\sum\limits_{{\bf k}, {\bf k}'} V({\bf k},{\bf k}') F({\bf k}' -
{{\bf q} / 2},{\bf k} + {{\bf q} / 2}) e^{i {\bf q} \cdot {\bf
r}}$, where  according to Eq.~(\ref{H}) the interaction $V({\bf
k},{\bf k}') = -\lambda_l \chi_l(\phi) \chi_l(\phi')$ and $F$ is
the standard Gor'kov's Green's function. In what follows we will
concentrate on the spatial dependence of the gap and assume that
its symmetry in the ${\bf k}$-space is preserved: $\Delta_{\bf
k}({\bf r}) = \Delta({\bf r}) \chi_l({\bf k}) \equiv \Delta_0
f({\bf r}) \chi_l({\bf k})$. Using these notations, we arrive at
the following free energy for the system expressed in terms of the
inhomogeneous order parameter $\Delta({\bf r})$
\begin{eqnarray}
\label{F} {\cal F}[\Delta,U] = && \!\!\!\!\!\!
\frac{1}{2}\int_{1,2} \Delta^*({\bf r}_1) A({\bf r}_1,{\bf r}_2)
\Delta ({\bf r}_2)\\
&&\!\!\!\!\!\!\!\!\!\!\!\! + \frac{1}{4}\int_{1,2,3,4}
\Delta^*({\bf r}_1) \Delta^*({\bf r}_2) B({\bf r}_1,{\bf r}_2,{\bf
r}_3,{\bf r}_4) \Delta({\bf r}_3)\Delta({\bf r}_4), \nonumber
\end{eqnarray}
 where $A({\bf r}_1,{\bf r}_2) = \lambda_l^{-1} \delta({\bf r}_1
- {\bf r}_2) - C({\bf r}_1,{\bf r}_2)$ and the Cooperon, ${\hat
C}$, is a random matrix expressed through the Green's functions
(before averaging over disorder) as follows $C({\bf r}_1,{\bf
r}_2) = T \sum\limits_{\varepsilon_n} G(\varepsilon_n;{\bf
r}_1,{\bf r}_2)G(-\varepsilon_n;{\bf r}_1,{\bf r}_2)$. As long as
we are interested in the location of the classical
(finite-temperature) phase transition, the dynamics of the order
parameter and the Cooperon are not important. We present the
Cooperon as a superposition of a local ``mean-field'' part and a
disorder dependent correction $\hat{C} = \langle{\hat{C}}\rangle +
\delta \hat{C}$. The ``mean-field'' part is diagrammatically
described by a simple Cooper bubble, without the disorder ladder
(see Fig.~1a). Any disorder vertex correction vanishes due to the
unusual symmetry of the gap.
The line where the average Ginzburg-Landau
coefficient  vanishes $\langle{A}\rangle = 0$ determines the
mean-field transition and leads to the well-known
Abrikosov-Gor'kov's equation~\cite{AG}
\begin{equation}
\label{AG} \ln{T_{p0} \over T_p} = \psi \left( {1 \over 2} + {1
\over 4 \pi T_p \tau} \right) - \psi \left( {1 \over 2} \right),
\end{equation}
where $T_{p0}$ is the pairing temperature without disorder and
$\tau$ is the scattering time.
 This equation implies that the pair-breaking effect of the
conventional disorder potential in an anisotropic gap
superconductor [i.e., $\int d \phi \chi(\phi) = 0$] is identical
to that of a time-reversal perturbation in an $s$-wave
superconductor.\cite{AIL} We reiterate that Eq.~(\ref{AG}) is a
result of the averaging over disorder in the sample. Below we
study mesoscopic corrections to this result, which qualitatively
can be interpreted as local changes in the scattering time
$\tau({\bf r})$ in Eq.~(\ref{AG}).

\begin{figure}
\centering
\includegraphics[width=3.3in]{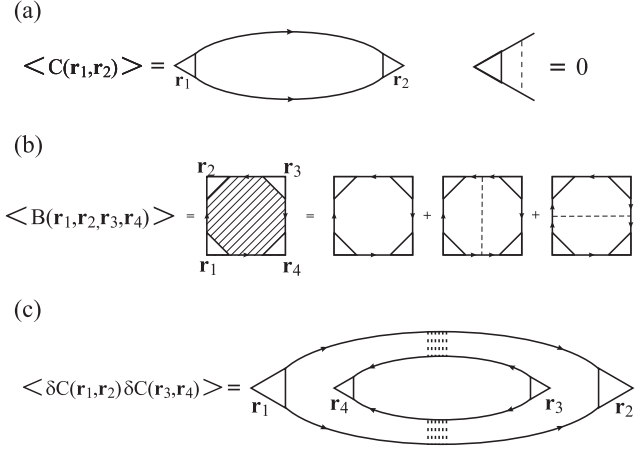}
\caption{\label{FIG:UCF}~(a)~The particle-particle bubble for an
anisotropic gap superconductor.  This Cooperon diagram contributes
to the coefficient in the quadratic term of the Ginzburg-Landau
expansion (\ref{F}). All disorder vertex corrections to the
Cooperon vanish.; (b)~Pictorial representation for the coefficient
in quartic term of the Ginzburg-Landau expansion (\ref{F}).
(c)~One of the UCF-type diagrams, which contribute to the
mesoscopic fluctuations of the transition point and mesoscopic gap
fluctuations.}
\end{figure}

We note that strictly speaking the nonlinear operator in the
quartic term of Eq.~(\ref{F}) is also random, however its
fluctuations can be neglected near the transition and its mean
field value can be used $\langle{B}\rangle$. This coefficient is
pictorially described by the Hikami-box diagram in Fig.~1b.  A
straightforward calculation of this diagram gives the following
general result:
\begin{eqnarray}
\label{B}  \langle{B}\rangle = -{\nu \over 16 \pi^2 T^2}
 \left[{ \alpha \over 12} \psi'''\left( {1 \over 2} + \alpha
\right)  + \overline{|\chi_l|^4}  \psi''\left( {1 \over 2} +
\alpha \right)  \right],
\end{eqnarray}
where $\nu$ is the density of states, $\alpha = (2 \pi T_p
\tau)^{-1}$, and the overline implies averaging over the Fermi
surface. In the clean limit, Eq.~(\ref{B}) reproduces the result
of Feder and Kallin~\cite{Kallin}, $\langle{B}\rangle = {7
\zeta(3) \nu / (8 \pi^2 T^2)}$. We note that the dirty limit is
not reasonable in the context of an anisotropic gap
superconductor, since it  implies that the pairing temperature is
suppressed to zero and there is no superconductivity. The maximum
impurity concentration which allows for superconductivity (the
quantum critical point) is $T_{p0} \tau_{\rm QPT} = \gamma /\pi
\sim 1$, where $\gamma \approx 1.781$ is the exponential of the
Euler's constant.


To find the local variations of the transition temperature, we
consider the following eigenvalue problem for the random matrix
$\delta \hat{C}$
\begin{equation}
\label{eigen} {1 \over \nu} \int d^2 {\bf r}' \delta C ({\bf
r},{\bf r}') \Delta ({\bf r}') = \epsilon \Delta({\bf r})
\end{equation}
and define the probability distribution function (PDF) of its
eigenvalues $\rho(\varepsilon) = \left\langle \delta \left(
\epsilon - \epsilon \left[ \delta \hat{C} \right] \right)
\right\rangle$. The averaging is performed over the PDF of the
random Cooperon matrix, which we assume Gaussian $P \left[ \delta
{\hat C} \right] \propto \exp \left[ - {1 \over 2} \delta {\hat C}
* \hat{\hat{K}}^{-1} * \delta {\hat C} \right]$, where the
asterisk implies a convolution over the two spatial variables and
the operator $\hat{\hat{K}}$ corresponds to the correlator of two
Cooperon operators, which in position representation has the form
$K({\bf r}_1,{\bf r}_2;{\bf r}_3,{\bf r}_4) = \left\langle \delta
C ({\bf r}_1,{\bf r}_2)\delta C ({\bf r}_3,{\bf
r}_4)\right\rangle$. This disorder-averaged correlator can be
calculated using the standard diagrammatic technique (see
Fig.~1c). These diagrams are topologically equivalent to the
universal conduction fluctuation (UCF) diagrams. However there are
important differences: (i)~First, here we are interested in the
Cooper channel and (ii)~Second, we are interested in the {\em
local} physics, not in a long-wavelength behavior of the
correlator.

Using the standard technique,\cite{NAL} we find the following
expression for the correlator
\begin{eqnarray}
\label{K1} K_1[ \{ {\bf r}_i \} ] = &&\!\!\!\! \delta\left({\bf
r}_1 - {\bf r}_4 \right) \delta\left({\bf r}_2 - {\bf r}_3 \right)
\left[ {2 \tau \overline{|\chi_l|^2} \over 4 \pi^2 D \nu }
\right]^2
\\ \nonumber && \!\!\!\!\!\!
\times {\int\limits_{\tau \to 0}^\infty}\int\limits_{\tau \to
0}^\infty {dt_1 dt_2 \over t_1 t_2 (t_1 + t_2)^2} \exp{\left[ -
{t_1 + t_2 \over 4 D t_1 t_2} \left| {\bf r}_1 - {\bf r}_3
\right|^2 \right]}.
\end{eqnarray}
Here the index ``1'' implies that we consider only one among all
possible UCF-type diagrams. However, they all contribute equally
to the correlator of interest and lead to a combinatorial factor
of $c$, which is equal to $c = 12$ in the orthogonal ensemble and
$c = 6$ in the unitary ensemble (e.g., in the presence of a
magnetic field). 

 The PDF of the local transition
temperatures and the corresponding gap amplitudes can be obtained
using the  optimal fluctuation method\cite{HL}
\begin{equation}
\label{PDF.Tc} \left\langle \rho(\epsilon) \right\rangle \propto
\exp\left[ - {1 \over 2} {\epsilon^2 \over \left\langle f \otimes
f\, \left| \hat{\hat{K}} \right| \, f \otimes f \right\rangle}
\right],
\end{equation} where the eigenvalue $\epsilon$ has the physical
meaning of a local pairing temperature fluctuation and $f({\bf
r})$ is a normalized function, which describes the spatial profile
and the shape of a single disorder-induced superconducting [if
$\epsilon > (T - \langle T_p \rangle)/T$] or metallic [if
$\epsilon < (T - \langle T_p \rangle)/T$] puddle. Strictly
speaking the latter function must be found from a non-linear
integral equation  $\epsilon f({\bf r}) = \Lambda \int_{1,2,3}
f^*({\bf r}_1) f^*({\bf r}_2) K({\bf r}_1,{\bf r}_2,{\bf r}_3,{\bf
r}) f({\bf r}_3) $ (where $\Lambda$ is a Lagrange multiplier which
appears in the optimal fluctuation method; see
Ref.~[\onlinecite{HL}] for technical details). However, one can
get a quantitatively reliable description of the PDF by
considering the puddle function to be a Gaussian of a
characteristic size $\xi$, i.e., $f(r) = \left( \pi \xi^2
\right)^{-1} \exp\left(-{r^2 \over 2 \xi^2}\right)$. In principle,
one can study the distribution of puddle shapes by decomposing the
function $f({\bf r})$ into spherical harmonics. We do not attempt
a study of the puddle shapes here, but just point out that
``higher-orbital momentum puddles'' are less probable than
spherically symmetric ones; the probability of finding a droplet
with ``momentum'' $m$ scales as $p_m\propto p_0^{m}$, where $p_0$
is the probability of a spherical puddle. To find the latter we
explicitly calculate the correlator in Eq.~(\ref{PDF.Tc}) and find
$\left\langle f \otimes f\, \left| \hat{\hat{K}} \right| \, f
\otimes f \right\rangle = {3 c \over 4 \pi} {l^2 \over \xi^2} {1
\over g^2}$, where  $\xi$ is the size of a puddle, $l$ is the mean
free path, and $g = E_{\rm F} \tau / \pi$ is the dimensionless
conductance. This leads to the following PDF of $T_p$'s
(\ref{PDF.Tc}):
\begin{equation}
\label{PDF.Tc.res} \left\langle \rho(\xi,T_p) \right\rangle
\propto \exp \left[ - {4 \pi \over 3 c} \left( {\xi \over l}
\right)^2 g^2 \left( {T_p - \langle T_p \rangle \over \langle T_p
\rangle} \right)^2 \right].
\end{equation}

\begin{figure}
\centering
\includegraphics[width=3in]{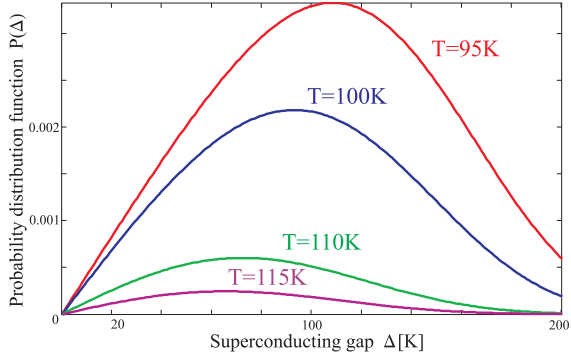}
\caption{\label{FIG:PDF}~(Color online) Plotted are distribution
functions of the superconducting gap $\Delta$ for various
temperatures, $T$. We assumed the following parameters: $g = 10$,
$E_F/T_p \sim 40$, $T_p = 93~K$. The latter choice is motivated by
the experimental work \protect{\cite{Yazdani}} on
Bi$_2$Sr$_2$CaCu$_2$O$_{8+\delta}$, where inhomogeneities have
been observed. The temperature behavior of the PDF following from
the mesoscopic fluctuation theory  is qualitatively similar to
that observed in experiment.\protect{\cite{Yazdani}} However, we
note here that since our approach is based on the BCS theory, the
theoretical results may not provide a quantitatively accurate
description of the cuprates.}\vspace*{-0.15in}
\end{figure}

We note that by applying a  magnetic field, one can cross over
from the orthogonal to the unitary ensemble and change the
combinatorial factor in (\ref{PDF.Tc.res}) from $c = 12$ to $c=6$,
which may be experimentally testable and should manifest itself as
a diminishing of the random $T_p(H)$ or gap distribution width
exactly by the factor of two. To find the PDF of the
superconducting gaps, we can just use the Ginzburg-Landau equation
(\ref{F}) and set $\Delta_0 = \sqrt{\nu \left( T_p - T \right) / (
\langle T_p \rangle \langle B \rangle )}$, where $\langle B
\rangle$ is given by Eq.~(\ref{B}). We note that it makes sense to
consider a finite-size droplet with a well-defined local
transition temperature or a gap, only if the size of the droplet
is much larger than the coherence length $\xi \gg
\xi_{\Delta_{0}}$. The opposite limit corresponds to the case of a
mesoscopic superconducting nanograin in which the notion of the
gap is not well-defined (see Ref.[\onlinecite{nano}] for a
review). Therefore, the smallest possible size of the puddle with
a well defined gap $\Delta_0$ is $\xi_{\rm min} \sim v_{\rm
F}/{\Delta_0}$, which implies that the dimensionless ratio
$\xi_{\rm min} /l \sim (T_p \tau)^{-1}$. The latter parameter is
of order one and thus the PDF of the gaps takes the form (see also
Fig.~\ref{FIG:PDF})
\begin{equation}
\label{PDF.gaps} P[\Delta] \sim {2 g \sqrt{\beta} b \Delta \over
\sqrt{\pi} \langle T_p \rangle^2} \exp \left[ - \beta g^2 \left(
{b \Delta^2 \over \langle T_p \rangle^2}  + {  T - \langle T_p
\rangle
 \over \langle T_p \rangle } \right)^2 \right],
\end{equation}
where in the $d$-wave case, $b \sim -\left[ {\alpha \over 3}
\psi'''\left({1 \over 2} + \alpha\right)  + 3 \psi''\left({1 \over
2} + \alpha\right) \right]/(32 \pi^2)$, $\alpha = (E_F/\langle T_p
\rangle) (2 \pi^2 g)^{-1}$, and $\beta \sim 1$. Note that in
Eq.~(\ref{PDF.gaps}) we have omitted a term proportional to
$\propto \delta(\Delta)$, which describes normal regions. The
physical picture, which emerges for such a disordered
superconductor is that right above the mean-field transition
temperature there exist rare superconducting islands separated by
``normal regions'' (which are still very close to the local
transition temperature).  We note that since the parameter $g
T_{p0}/E_{\rm F}$ is at best of order one, the ``mesoscopic
Griffiths phase'' overlaps with the Ginzburg region of strong
superconducting fluctuations.\cite{LV} This leads to the
conclusion that the Griffiths phase is a mixture of
superconducting islands and metallic regions with strongly
suppressed density of states.

\begin{figure}
\centering
\includegraphics[width=3in]{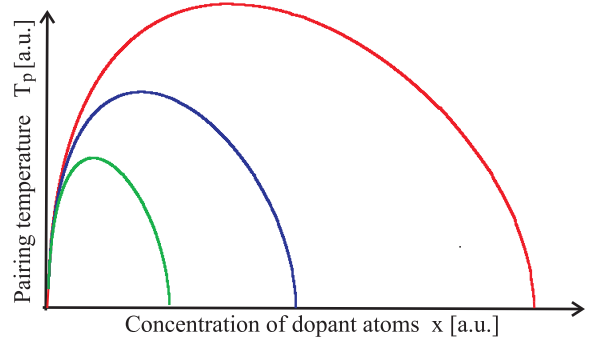}
\caption{\label{FIG:dome}~(Color online) This figure is to
illustrate the qualitative discussion in the text about the
possibility of a dome-shaped doping dependence of the pairing
temperature within the Abrokosov-Gor'kov theory. The figure shows
a typical dependence of $T_p$ on ``doping,'' x, in the toy model
of the Abrikosov-Gor'kov theory with both $T_{p0}(x)$ and
$\tau(x)$ being dependent on the same doping parameter. The graphs
are solutions of the following equation for $t_p$:
$\ln{\left(x^{1-\alpha}/t_p\right)} = \psi\left(1/2 + x
\delta/t_p\right) - \psi(1/2)$, where $t_p$ is a dimensionless
pairing temperature and $\delta$ characterizes the strength of
disorder. The graphs correspond to $\alpha=1/2$ and $\delta = 0.1$
(red line), $0.15$ (blue line), and $0.2$ (green line).}
\end{figure}

Even though our quantitative description directly applies only to
a weakly coupled BCS superconductor, we believe that some aspects
 of the theory are relevant to the
cuprates as well ( the importance of disorder effects for the
cuprates have been discussed previously, see, e.g.
[\onlinecite{Engel,H1,H3,H4,Balatsky,Maska}]). But first, we make
the following curious observation: In a high-$T_c$ superconductor
the major source of disorder is presumably the dopant  atoms. But
this access oxygen is also the source of the carriers, which lead
to superconductivity in the first place. Thus, the ``clean''
pairing temperature in the Abrikosov-Gor'kov's formula (\ref{AG})
should explicitly depend on the doping level, $T_{p0}(x)$ (e.g.
through a BCS-like high-energy cut-off or possibly via a locally
modulated electron pairing interaction.\cite{H4}) The scattering
time depends on $x$ too and contributes to superconductivity
suppression. Thus there are two competing effects of the dopants:
They both enhance and suppress superconducting properties. An
interesting result, which follows from the simple
Abrikosov-Gor'kov's equation (\ref{AG}), is that even if
$T_{p0}(x)$ is monotonically increasing with $x$, but $T_{p0}(x)
\tau(x) \propto x^{-\alpha}$ with $\alpha > 0$, then the actual
pairing temperature doping dependence has a dome-shaped form. An
example of such a dependence for $\alpha =1/2$  is plotted in
Fig.~\ref{FIG:dome}.

The mesoscopic disorder fluctuations in the Abrikosov-Gor'kov
theory may not correspond to the modulations of the transition
temperature in a high-$T_c$ superconductor, but should be related
to the modulations of the pseudogap temperature, $T_*$, which is
believed to be the onset of Cooper pairing (i.e., to $T_p$, but
not $T_c$). The ``pseudogap'' region $T_c<T<T_p$ presumably
represents the regime of strong phase fluctuations~\cite{Kivelson}
and the superconducting transition is expected to be that of
$XY$-type. In the latter scenario, the transition temperature is
proportional to the superfluid density, which in turn is directly
related to the local value of the gap. The mean field gap is
determined by the deviation $(T_p - T)$ from the local pairing
temperature (e.g., via a non-linear BCS-like self-consistency
equation $\delta S/\delta \Delta = 0$). If $T_p({\bf r})$
fluctuates due to disorder, so does $\Delta({\bf r})$ and, quite
generally, the local gap should ``follow''  local $T_p$. In fact,
such a correlation has been observed in experiment.\cite{Yazdani}
In the model of uncorrelated short range disorder, the only
possible result for the corresponding PDF is Eq.~(\ref{PDF.gaps}),
with $\Delta$ centered in the vicinity of the mean-field gap at
the given temperature $P[\Delta] \propto \exp\left\{ - \beta g^2
\left[ 1 -\Delta/ \langle \Delta (T) \rangle \right]^2 \right\}$.
In a clean material, the corresponding regime of mesoscopic
fluctuations is very narrow unless there are other types of
disorder effects, such as ``structural disorder'' (e.g., extended
defects, warping of the 2D planes, etc.), which can be modelled as
a random diffusion coefficient~\cite{GaL} $\left\langle D({\bf
r})D({\bf 0}) \right\rangle =  \langle D \rangle^2 a^2 \delta({\bf
r})$.  These phenomena lead to qualitatively the same effect of
random $T_p$ and $\Delta$, but, may occur at the length-scales
much larger than the mean-free path and become important in a much
wider range of parameters [the width of the distribution is
determined by $ (a/l) g^{-1}$ instead of $g^{-1}$]. We also note
that if indeed the local gap fluctuations observed in
experiment~\cite{Yazdani} are related to mesoscopic disorder
effects, they should be correlated with the pinning
properties,\cite{LO} i.e., the width of the gap distribution
should be proportional to the critical current $j_c$ in the
collective pinning regime~\cite{GaL}, $\left\langle \left(1 -
\Delta /\langle\Delta\rangle \right)^2 \right\rangle \propto
(j_c/j_{c0}) (H/H_{c2})$ (where $j_{c0}$ is the critical current
in zero field).

The author acknowledges the Aspen Center for Physics for
hospitality and the JQI for financial support.

\bibliography{disorder}

\end{document}